# Spin-1/2 one- and two- particle systems in physical space without *eigen*-algebra or tensor product

Sokol Andoni*, Technical University of Denmark, Dept. of Chemistry, Kgs. Lyngby 2800, Denmark

*Corresponding author, sond4p@gmail.com

**Abstract.** Under the spin-position decoupling approximation, a vector with a phase in 3D orientation space endowed with geometric algebra, substitutes the vector-matrix spin model built on the Pauli spin operator. The standard quantum operator-state spin formalism is replaced with vectors transforming by proper and improper rotations in the same 3D space – isomorphic to the space of Pauli matrices. In the single spin case the novel spin ½ representation: (1) is Hermitian; (2) shows handedness; (3) yields all the standard results and its modulus equals the total spin angular momentum $S_{tot} = \sqrt{3}\hbar/2$; (4) formalizes *irreversibility* in measurement; (5) permits adaptive imbedding of the 2D spin space in 3D. Maximally entangled spin pairs (1) are *in phase* and have *opposite* handedness; (2) relate by one of the four basic *improper* rotations in 3D: plane-reflections (triplets) and inversion (singlet); (3) yield the standard total angular momentum; (4) all standard expectation values for bipartite and partial observations follow. Depending on whether proper and improper rotors act one – or two – sided, the formalism appears in two complementary forms, the 'spinor' or the 'vector' form, respectively. The proposed scheme provides a clear geometric picture of spin correlations and transformations entirely in the 3D physical orientation space.

## 1. Introduction

Key developments in Quantum Mechanics (QM), such as the first phenomenological description of spin ½ by Pauli [1] and the first quantum relativistic description of the electron by Dirac [2], 're-invented' Clifford algebras (in matrix representation), seemingly unaware of Grassmann's [3] and Clifford's [4] works more than half a century earlier. The promotion of vector-generated Clifford algebras in physics and in particular the development of the spacetime algebra (STA) was undertaken by Hestenes [5, 6], with more scientists joining in during the last 2-3 decades [7-10]. Spin formalism is arguably the main application of STA in QM [7]. In the STA literature [6, 7] it has become customary to represent spin by a bivector or the spin vector normal to it, both of modulus $\hbar/2$. Notably, such representations of spin with quantum number $s = 1/2$ *do not* comprise a quantity corresponding to the ubiquitous standard Pauli vector spin operator $\hat{\boldsymbol{\sigma}} = (\hbar/2)\hat{\sigma}_j \mathbf{x}_j$, with $\hat{\sigma}_j$ the Pauli matrices and $\mathbf{x}_j$ the unit 3D frame vectors. The *total* quantum spin angular momentum is directly related to the modulus of $\hat{\boldsymbol{\sigma}}$, $S_{tot}^2 = |\hat{\boldsymbol{\sigma}}|^2 = 3\hbar^2/4 = \hbar^2 s(s+1)$, and it is three times the square of the *observed* spin angular momentum along, let say $\mathbf{x}_3$, $S_3^2 = (\hbar|\hat{\sigma}_3|/2)^2 = \hbar^2/4$. According to the standard QM interpretation [11], when the $S_3$ component is well-defined by measurement, the other two components of spin are not zero, but fluctuate between $+\hbar/2$ and $-\hbar/2$, therefore $S_{tot}^2 = 3S_3^2$. This interpretation builds on the uncertainty principle as expressed by the mutual non-commutativity of the Pauli matrices $\hat{\sigma}_j$.

Recently I generalized STA – a 16D Clifford algebra $\mathcal{Cl}_{(1,3)}$ with signature $(+ - - -)$ on a 4D real vector space, to spacetime - reflection (STR) – a 32D $\mathcal{Cl}_{(2,3)}$ on a 5D real vector space [12]. The equivalent real dimension of the space of Dirac matrices is also 32. The action of the geometric product onto the quintet of orthonormal spacetime-reflection frame vectors $\{e_\mu, e_5\}$ generates the algebra. The *reflector* $e_5$ is Hermitian and the suffix 5 (instead of 4) emphasizes the analogy with the standard Dirac $\gamma_5$ matrix. With the geometric pseudoscalar of STR $\dot{\mathrm{I}} \equiv e_0 e_5 e_1 e_2 e_3 \equiv e_{05123}$ (see Eq. (3) below), the frame-free Dirac equation (DE) in STR, $\hat{p}\psi = \dot{\mathrm{I}}\hbar\nabla\psi = m\psi$ $(c = 1)$, follows from the direct quantization of the relativistic 4-momentum *vector* with modulus equal to the rest energy. In the nonrelativistic regime, STR DE gives rise to the STR Pauli equation, STR PE, which has the same form as the standard PE. The interested reader can find all this and more in ref. [12]. For a brief description, see the Appendix.

The main motivation for the present work has been to model spin with the help of geometric algebra, but from a *different perspective* compared to the STR PE. Physically, in the spin-position decoupling regime, spin ½ 'lives' in the 3D orientation space, therefore it is relevant to model it in this space. The model presented in the following is inspired by the *vector* model of angular momentum, in particular that of spin based on the Pauli matrix-vector operator $\hat{\boldsymbol{\sigma}}$ [11] mentioned above. In Section 3, I recast the standard Pauli *operator* into the 3D orientation subspace of STR, obtaining a *vector* model with a phase to represent spin. We can reproduce all standard results for one & two spin ½ systems; also, like in the case of $\hat{\boldsymbol{\sigma}}$ the full spin modulus equals $S_{tot}$. At the same time the model helps visualize how the 2D spin space, where orthogonality relations apply, can be imbedded adaptively in 3D space. All this follows without needing the *eigen*-algebra and the *tensor* product, which seem so indispensable in standard QM. The simple reason is that we essentially work here with vectors in 3D and the standard operator – eigenstate/ eigenvalue formalism is absent in the present scheme, thus naturally allowing different spins to belong to the same 3D space, as they physically do. As a result, e.g. zero total spin in the singlet state can be expressed by the sum of two opposite vectors in 3D, without producing a dull zero-state as in the standard formalism (if it were not for the tensor product). Superposition hinders the entangled pair to appear as two separate single spins. The pair can have any direction in space, thus embodying the spherical symmetry of the singlet state.

We present a *'kinematic' model* of spin. In the nonrelativistic regime all dynamics is still governed by the STR PE [12]. The 'extrinsic', phase-insensitive part of the present model is equal to the spin vector, thus making *contact* to the STR PE, the STA or the observables in the standard formalism. On the other hand, the phase sensitive part, which in measurements produces zero expectation values, contributes to the 'intrinsic' full spin modulus $S_{\text{tot}}$ and hints to a hidden structure of spin, like in standard QM and unlike in STA. As mentioned above, a well-defined spin along one axis, leaves uncertain and with zero expectation value the two components orthogonal to it. This interpretation builds on the non-commutativity of the Pauli matrices. Accordingly, spin in the present model appears as a *sum* of three (anti-commuting) 3D frame vectors, instead of e.g. a commuting vector – bivector pair. These statements will become clearer in the following.

The rest of the report comprises three Sections. In Sec. 2, I introduce basic concepts of geometric algebra as well as the bases of STR and two of its subspaces. The new definition of spin opens Sec. 3. Transformations for the one and two particle cases appear in it in *vector* form, i.e. as *two-sided* rotor/ reflection operations. In Sec. 4 on the *spinor* form, transformations for the same two cases appear as *one-sided* operations. Orthogonality relations, e.g. between spin up and spin down hold in the last representation. The connection between the vector and the spinor forms is also discussed in Sec. 4 followed by the Conclusions. The added Appendix is a succinct presentation of STR DE, STR PE and the corresponding forms of the STR spinors.

**2. A Short Introduction to Geometric Algebra**

The geometric or Clifford product of three vectors $\mathbf{u}, \mathbf{v}, \mathbf{w}$, for example in a 3D Euclidean space, combines Hamilton's scalar (symmetric) and Grassmann's wedge (antisymmetric) products; if not zero it is invertible:

$$\mathbf{uv} = \mathbf{u} \cdot \mathbf{v} + \mathbf{u} \wedge \mathbf{v} = \mathbf{v} \cdot \mathbf{u} - \mathbf{v} \wedge \mathbf{u}; \quad (\mathbf{uv})\mathbf{w} = \mathbf{u}(\mathbf{vw}) = \mathbf{uvw}; \quad (\mathbf{uv})^{-1} = \mathbf{vu}/\mathbf{u}^2\mathbf{v}^2. \tag{1}$$

In terms of anti-commutator and commutator brackets:

$$\{\mathbf{u}, \mathbf{v}\} \equiv \mathbf{uv} + \mathbf{vu} = 2\mathbf{u} \cdot \mathbf{v}; \quad [\mathbf{u}, \mathbf{v}] \equiv \mathbf{uv} - \mathbf{vu} = 2\mathbf{u} \wedge \mathbf{v}. \tag{2}$$

The geometric product generalizes to any dimension and space signature, whereas the cross product is valid only in 3D. For $\mathbf{u}, \mathbf{v} \in \boldsymbol{\Sigma}$ (see Eq. (5) below) the two relate by $\dot{I}(\mathbf{u} \times \mathbf{v}) = \mathbf{u} \wedge \mathbf{v} = \frac{1}{2}[\mathbf{u}, \mathbf{v}]$ (a bivector).

As mentioned in the Introduction, spacetime-reflection, STR generalizes STA. The pseudoscalar in STR is the pentavector $e_{05123} \equiv \dot{I}$, which commutes with all elements of STR and a basis for the STR $\mathcal{Cl}_{(2,3)}$ is:

$$\{1, e_\tau, e_\tau \wedge e_\upsilon, \dot{I}(e_\tau \wedge e_\upsilon), \dot{I}e_\tau, \dot{I}; \tau, \upsilon = 0,1,2,3,5; \tau \neq \upsilon; \text{signature } \zeta_{\tau\upsilon} \equiv e_\tau \cdot e_\upsilon = (+ - - - +)\delta_{\tau\upsilon}\}. \quad (3)$$

(3) is the union of the basis of STA (generated by $\{e_\mu\}$) and its product with $e_5$. Two 3D subspaces of relative vectors (bold upright symbols) in STR will be of interest here:

$$\mathbf{X}: \{1, \mathbf{x}_j \equiv e_{0j}, \mathbf{x}_j \mathbf{x}_k \equiv \mathbf{x}_{jk}, \mathbf{x}_{123}; \ j, k = 1,2,3; \ j \neq k\}; \text{ generated by the boost (polar) vectors } \{\mathbf{x}_j\} \quad (4)$$

and

$$\boldsymbol{\Sigma}: \{1, \boldsymbol{\sigma}_j \equiv e_{0j5}, \boldsymbol{\sigma}_{jk} = \mathbf{x}_{jk} = \epsilon_{jkl}\dot{I}\boldsymbol{\sigma}_l, \boldsymbol{\sigma}_{123} = \dot{I}; \ j \neq k\}; \text{ generated by the spin (axial) vectors } \{\boldsymbol{\sigma}_j\}. \quad (5)$$

The orientation space with frame vectors $\{\boldsymbol{\sigma}_j\}$ and algebra $\mathcal{Cl}_{(3,0)}$ given by $\boldsymbol{\Sigma}$ is an example of 3D space where the cross product mentioned above is valid. The appropriate form of pseudoscalar in this case is the trivector in (5) depicted by the same symbol $\dot{I}$ as the pentavector in (3); the last is obtained from the first by substituting $\boldsymbol{\sigma}_j$ with $e_{0j5}$. $\mathbf{X}$ corresponds in STR to the even subspace of STA. It is appropriate in STR to use the notation $\boldsymbol{\sigma}_j$ for spin vectors, which are *axial* vectors, i.e. they are parity-even, $e_0 \boldsymbol{\sigma}_j e_0 = \boldsymbol{\sigma}_j$, while the $e_0 \mathbf{x}_j e_0 = -\mathbf{x}_j$ of STA are parity-odd. The intersection of $\mathbf{X}$ with $\boldsymbol{\Sigma}$: $\mathbf{X} \cap \boldsymbol{\Sigma} = \{\mathbf{x}_j \mathbf{x}_k \equiv \mathbf{x}_{jk} = \boldsymbol{\sigma}_{jk} = \delta_{jk} + \varepsilon_{jkl}\dot{I}\boldsymbol{\sigma}_l\}$ consists of the real scalar and the bivectors. Here I address spin in the non-relativistic regime under the spin–position decoupling approximation [6, 7, 11-13]. The 3D physical space in this case reduces to *orientation space* at a point (the origin) and the relevant symmetry operations are proper rotations and reflections. In STR it corresponds to the subspace $\boldsymbol{\Sigma}$ in (5) [12] with a Clifford algebra $\mathcal{Cl}_{(3,0)}$ isomorphic to that of Pauli matrices, therefore the notation [9, 12]. Although $\boldsymbol{\Sigma}$ differs from $\mathbf{X}$, as discussed above, $\boldsymbol{\Sigma}$ in STR and the even subspace in STA have the same dimension and isomorphic algebra $\mathcal{Cl}_{3,0}$ [7, 12]. Geometrically, the bivectors $\dot{I}\boldsymbol{\sigma}_l$, e.g. $\dot{I}\boldsymbol{\sigma}_2 = \boldsymbol{\sigma}_{31}$, represent oriented surface elements, while the pseudoscalar $\dot{I} = \boldsymbol{\sigma}_{123}$ is an oriented volume element, all unitless. A rotor $R_\vartheta$ with axis along the unit vector $\mathbf{u}$ is a *unitary* transformation in $\boldsymbol{\Sigma}$ (length, angle and handedness preserving):

$$\Sigma \ni A \rightarrow R_\vartheta A R_\vartheta^\dagger \equiv e^{-i\mathbf{u}\vartheta/2} A e^{-i\mathbf{u}\vartheta/2} \in \Sigma; \; R_\vartheta = \cos\frac{\vartheta}{2} - i\mathbf{u}\sin\frac{\vartheta}{2}. \tag{6}$$

Rotors can alter vectors and bivectors in $\Sigma$. Inversion and plane-reflections are another type of (non-unitary) transformations, which also conserve lengths and angles, but *invert* handedness, e.g. of the triplet $(\sigma_1, \sigma_2, \sigma_3) \equiv \{\sigma_j\}$. The three frame reflections and the inversion can be depicted by ($\sigma_0 = 1$; $\mu = 0,1,2,3$):

$$\Sigma \ni A \rightarrow i\sigma_\mu A i\sigma_\mu = -\sigma_\mu A \sigma_\mu \in \Sigma; \; \{\sigma_j\} \rightarrow i\sigma_\mu \{\sigma_j\} i\sigma_\mu = (-1)^{\delta_{\mu 0}}\{(-1)^{\delta_{\mu j}}\sigma_j\} \in \Sigma. \tag{7}$$

It is now appropriate to present the novel definition of spin 1/2.

## 3. Definition and vector (two-sided) transformations of spin 1/2 for one and two particle systems.

As mentioned in the Introduction, the present model is inspired by the heuristic vector model of spin in the standard formalism [11]. The orientation space we work in here is the subspace $\Sigma$ of STR presented in (5), which is isomorphic to the space of Pauli matrices. One way to remake in $\Sigma$ the standard Pauli vector spin operator $(2/\hbar)\hat{\boldsymbol{\sigma}} = \hat{\sigma}_j \mathbf{x}_j = (\hat{\sigma}_1 \mathbf{x}_1 + \hat{\sigma}_2 \mathbf{x}_2)(\text{off} - \text{diagonal}) + \hat{\sigma}_3 \mathbf{x}_3 (\text{diagonal})$ (with $\hat{\sigma}_j$ the Pauli matrices and $\mathbf{x}_j$ the unit frame vectors in 3D), is to take the sum of the three frame vectors in $\Sigma$ and add a phase $\varphi$ rotating $\pm\sigma_1 + \sigma_2$ around $\pm\sigma_3$. By this we remake a symbolic 'vector' operator having matrix components into a *vector with a phase*; $\pm\sigma_1 + \sigma_2$ (with phase) correspond to the off-diagonal terms in $\hat{\boldsymbol{\sigma}}$ and $\pm\sigma_3$ to the diagonal terms. $\pm\sigma_3$ define, as by usual convention, the spin vectors up and down in the novel model (see Definition (8)). By remaking the 'master' Pauli operator $\hat{\boldsymbol{\sigma}}$ (with components in the Hilbert space and with unit vectors in 3D space) into a couple of spins $S_\sigma$ in $\Sigma$ we also *loose* the standard operator-state formalism; instead, we have now vectors in $\Sigma$ transforming by the operations of proper and improper rotations.

**3.1. *One-particle systems.*** More precisely, the spin up $\uparrow_\sigma$ (down $\downarrow_\sigma$) along the $\sigma_3$ axis is defined by the sum of three orthonormal vectors scaled by $\hbar/2$ together with a two-sided rotor $R_\varphi = e^{-i\sigma_3 \varphi/2}$ in the plane $\sigma_{12}$:

$$S_{\uparrow\sigma} = \frac{\hbar}{2} R_\varphi (\sigma_3 + \sigma_1 + \sigma_2) R_\varphi^\dagger = \frac{\hbar}{2}(\sigma_3 + R_\varphi(\sigma_1 + \sigma_2)R_\varphi^\dagger) \equiv \frac{\hbar}{2}(\sigma_3 + (\sigma_1 + \sigma_2)_\varphi);$$

$$S_{\downarrow\sigma} = \sigma_2 S_{\uparrow\sigma} \sigma_2 = \frac{\hbar}{2}(-\sigma_3 + (-\sigma_1 + \sigma_2)_{-\varphi}); \quad S_\uparrow S_\uparrow = S_\downarrow S_\downarrow = S_\sigma^2 = \frac{3\hbar^2}{4}; \quad S_\sigma^\dagger = S_\sigma; \; S_{-\sigma} \neq -S_\sigma;$$

$$\langle S_\sigma \rangle \equiv \frac{\hbar}{2}\left(\pm\sigma_3 + \langle R_{\pm\varphi}^2 \rangle(\pm\sigma_1 + \sigma_2)\right) \equiv \frac{\hbar}{2}\left(\pm\sigma_3 + (\langle\cos\varphi\rangle \mp \langle\sin\varphi\rangle i\sigma_3)(\pm\sigma_1 + \sigma_2)\right) = \pm\frac{\hbar}{2}\sigma_3. \tag{8}$$

Angled brackets alone $\langle \; \rangle$ denote the expectation value, while $\langle \; \rangle_0$ as in STA extracts the scalar part of an expression. The phase $\varphi$ degree of freedom is unobservable, which I express here by the zero expectation values $\langle\sin\varphi\rangle = \langle\cos\varphi\rangle = 0$ (see last equation in (8)). In other words, *measurement* extracts the *phase-* *insensitive* part of spin or combination of spins. The squared modulus $S_\sigma^2 = 3\hbar^2/4$ equals the squared QM total angular momentum of spin; it comprises equal contributions from the three components, exceeding by a factor of 3 the observed *squared spin modulus* $\langle S_\sigma \rangle^2 = \hbar^2/4$. Therefore, from the perspective of the present model, when we measure a definite value of spin, let say on the $\boldsymbol{\sigma}_3$ axis, the other spin components on the plane $\mathbf{i}\boldsymbol{\sigma}_3$ are *not* zero, but have a zero expectation value. This corresponds to the standard *uncertainty* on the spin components orthogonal to the measurement axis [11] as expressed by the pairwise noncommutativity of the spin components. Such a 'hidden' structure of spin is absent in STA.

*Handedness* is assigned from the vector triplet (with signs as) in the expression for spin, so that $S_\sigma$ (i.e. $S_{\uparrow\sigma}$ or $S_{\downarrow\sigma}$) in (8) are both right-handed ($r$), while inverted and plane-reflected $\mathbf{i}\boldsymbol{\sigma}_\mu S_\sigma \mathbf{i}\boldsymbol{\sigma}_\mu$ (see (7)) are all left-handed ($\ell$). The two subspaces $\{\ell\}$ and $\{r\}$ are *disjoint* under proper rotations (unitary transformations), while *reflections* and *inversion* (improper rotations) convert a spin from one subspace to the other. Any left-handed ($\ell$) spin-up (-down) is the *inverse* of a right-handed ($r$) spin-down (-up), which clearly demonstrates the linear dependence between the two in 3D:

$$S_{\sigma(\ell)} = \mathbf{i} S_{-\sigma(r)} \mathbf{i} = -S_{-\sigma(r)} \tag{9}$$

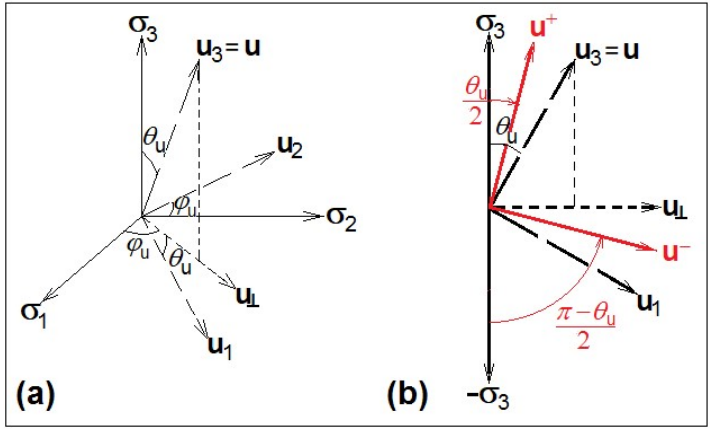

*Figure 1. Relation between the reference spin with spin vector $\boldsymbol{\sigma}_3$ and an arbitrary spin, represented by the spin vector $\mathbf{u}=\mathbf{u}_3$. Rendering of (a) the vector (two-sided rotor) transformation in 3D, see eq. (10) for the full spin and Eq. (12) for the spin vector alone; (b) reduced spinor ($\mathbf{u}^+$, $\mathbf{u}^-$) (one-sided rotor) transformation in the plane defined by $\boldsymbol{\sigma}_3$ and $\mathbf{u}$, see Eqs. (12), (33). The panel (b) shows the plane defined by $\boldsymbol{\sigma}_3$ and $\mathbf{u}_\perp$ or by $\mathbf{u}$ and $\mathbf{u}_1$ in panel (a).*

We therefore choose + in linear relations between spins in order to keep track of the handedness (see e.g. (34)). The reflection $\boldsymbol{\sigma}_2 S_\sigma \boldsymbol{\sigma}_2$ onto $\boldsymbol{\sigma}_2$ is equivalent in $\Sigma$ to a $\pi$-rotation of $S_\sigma$ around $\boldsymbol{\sigma}_2$. Focusing on $\{r\}$, any spin $S_u$ can be expressed by $S_\sigma$ and a combined rotation $R_u \equiv R_{\theta_u} R_{\varphi_u}$, with

$$R_{\theta_u} = e^{-\mathbf{i}\mathbf{u}_2 \theta_u/2}; \qquad -\mathbf{i}\mathbf{u}_2 = \mathbf{u}^-\mathbf{u}^+ =$$

$$\mathbf{u}_\perp \boldsymbol{\sigma}_3 = \mathbf{u}_{13} \quad \text{(see (6) and Figure 1a, b):}$$

$$(S_u)_\varphi = R_u (S_\sigma)_\varphi R_u^\dagger =$$

$$\tfrac{\hbar}{2} R_u (\boldsymbol{\sigma}_3 + (\boldsymbol{\sigma}_1 + \boldsymbol{\sigma}_2)_\varphi) R_u^\dagger = \tfrac{\hbar}{2} R_{\theta_u} (\boldsymbol{\sigma}_3 + (\mathbf{u}_\perp + \mathbf{u}_2)_\varphi) R_{\theta_u}^\dagger = \tfrac{\hbar}{2}[\mathbf{u} + (\mathbf{u}_1 + \mathbf{u}_2)_\varphi]. \tag{10}$$

Observable quantities here, as in the standard spin formalism, do not depend on handedness or on phase for both the one- and the two- spin cases. A *measurement* of $(S_\sigma)_\varphi$ by a Stern-Gerlach (*SG*) magnet aligned along $\mathbf{u} = \mathbf{u}_3$, transforms spin with complete *loss* of phase correlation, as shown schematically by:

$$(S_\sigma)_\varphi \xrightarrow{SG} (S_u)_{\varphi'}; \quad \langle \sin\varphi' \sin\varphi \rangle_{SG} = \langle \sin\varphi' \rangle \langle \sin\varphi \rangle = 0. \tag{11}$$

Then it is sufficient to restrict the transformation to only the *phase-insensitive* spin vector:

$$\tfrac{\hbar}{2}\boldsymbol{\sigma}_3 \to \tfrac{\hbar}{2} R_{\theta_u} \boldsymbol{\sigma}_3 R_{\theta_u}^\dagger = \tfrac{\hbar}{2}\left(\mathbf{u}^+ \cos\tfrac{\theta_u}{2} + \mathbf{u}^- \sin\tfrac{\theta_u}{2}\right) = \tfrac{\hbar}{2}\mathbf{u}, \text{ where: } R_{\theta_u}\boldsymbol{\sigma}_3 = \mathbf{u}^+; \; R_{\pi-\theta_u}^\dagger(-\boldsymbol{\sigma}_3) = \mathbf{u}^-. \tag{12}$$

Within a normalization factor, the two first expressions of spin in (12) are identical to the definition of spin vector in STA [7]. The corresponding expression for the full spin model (8) is the first equation in (10). It is also clear from Fig. 1a, b that *any* unit spin vector $\mathbf{u}$ can be expressed by means of the two unit *basis* spins $\pm\boldsymbol{\sigma}_3$ through the 'tailor-made' linear combination of the orthonormal 'spinor' pair $\mathbf{u}^+, \mathbf{u}^-$. This uniquely defined geometric construction illustrates the working of the quantum spin basis in 3D. Obviously, with only the spin vector involved, as in (12), there are no restrictions on handedness. The halfway vectors $\mathbf{u}^+, \mathbf{u}^-$ in Fig. 1b depict one form of *reduced* spinor representation and as shown by the last two equalities in (12) they arise from the action of one-sided rotors onto the spin vectors up and down. Of course, one can obtain $\mathbf{u}^+$ and $\mathbf{u}^-$ by one-sided action of rotors on spin-up alone; even then, *two* terms are needed with distinct probability amplitudes for coincidence and anti-coincidence ($\pm\boldsymbol{\sigma}_3$ are linearly dependent; $\mathbf{u}^\pm$ are not). For the sake of definiteness, we insist that a spin basis consist of two *opposite spins*. Eqs. (10-12) express the 'kinematics' of spin measurement, the last rendering explicit the *probability amplitudes* for coincidence & anticoincidence.

An important last remark about the spinor duplet $\mathbf{u}^+, \mathbf{u}^-$ is that it illustrates the geometry of *half-angles* characterizing quantum spin, here *flexibly* connecting the fixed opposite spins $\pm\boldsymbol{\sigma}_3$ with any new spin vector $\mathbf{u}$. Had we wished to rigidly anchor the spinor representation to the frame $\{\boldsymbol{\sigma}_3, \boldsymbol{\sigma}_1, \boldsymbol{\sigma}_2\}$ in the way it is done in the standard formulation or in STA, then we could have chosen e.g. $\theta_u = \varphi_u = 0$, see Fig. 1a. Then $\mathbf{u}^+ \to \boldsymbol{\sigma}_3$ and $\mathbf{u}^- \to \boldsymbol{\sigma}_1$, which reproduces the STA representation of spin up and spin down relative to $\boldsymbol{\sigma}_3, -\boldsymbol{\sigma}_3$ i.e. $\boldsymbol{\sigma}_3\{1, -i\boldsymbol{\sigma}_2\}$ (see also [12]). The same conclusion follows even more directly by using $R_{\theta_u}$ and $R_{\pi-\theta_u}^\dagger$ as representatives of spin up and spin down, respectively. It is clear from the last two relations in (12) that for fixed $\boldsymbol{\sigma}_3$ there is a one to one correspondence:

$$\{\mathbf{u}^+, \mathbf{u}^-\} \leftrightarrow \{R_{\theta_u}, -R_{\pi-\theta_u}^\dagger\}; \text{ in the limit } \theta_u = \varphi_u = 0 \text{ we obtain } \{R_{\theta_u}, -R_{\pi-\theta_u}^\dagger\} \to \{1, -i\boldsymbol{\sigma}_2\}, \tag{13}$$

which is precisely the STA representation for spin up and spin down. With the rigid choice, one would have to use also the half-angle $\varphi_u/2$ in the expression of the new spin with the help of the spin basis, just as it is the case in the standard formalism; see the rotor $R_u$ in Eq. (10). The flexibility and the geometric clarity of the present spinor construction in (12) and in Fig. 1b are certainly attractive.

After the Definition of spin in (8) and the demonstration in Equations (12, 13) and in Fig. 1 of how the pair of spins $\langle S_\sigma \rangle$ constitutes a spin basis in 3D (for the full spin see Eqs. (34-36)), it is time to show the correspondence with basic relations from the standard formalism. First, we generalize Definition (8) to spins $S_{\sigma j}$ with spin vectors $\boldsymbol{\sigma}_j$ ($j = 1,2,3$). Depicting for short $S_{\sigma j}$ by $S_j$ and with $\left(\boldsymbol{\sigma}_{j+1} + \boldsymbol{\sigma}_{j+2}\right)_{\varphi(j)} \equiv R_{\varphi(j)}(\boldsymbol{\sigma}_{j+1} + \boldsymbol{\sigma}_{j+2})R^\dagger_{\varphi(j)} = e^{-i\boldsymbol{\sigma}_j \varphi(j)/2}(\boldsymbol{\sigma}_{j+1} + \boldsymbol{\sigma}_{j+2})e^{i\boldsymbol{\sigma}_j \varphi(j)/2}$ (the indices $j+1, j+2$ are mod3), we adapt (8) in relation to each frame vector:

$$S_j \equiv \frac{\hbar}{2}\left(\boldsymbol{\sigma}_j + (\boldsymbol{\sigma}_{j+1} + \boldsymbol{\sigma}_{j+2})_{\varphi(j)}\right), \text{ e.g. } S_2 \equiv \frac{\hbar}{2}(\boldsymbol{\sigma}_2 + (\boldsymbol{\sigma}_3 + \boldsymbol{\sigma}_1)_{\varphi(2)}); \quad \langle S_j \rangle \equiv \frac{\hbar}{2}\boldsymbol{\sigma}_j. \tag{14}$$

It is clear from (14) that the $\langle S_j \rangle$ satisfy the same algebra as the Pauli matrices, as they should in order to represent the observed angular momentum of spin. In commutator form, this appears as follows:

$$\tfrac{1}{2}[\langle S_j \rangle, \langle S_k \rangle] = \varepsilon_{jkl} \tfrac{\hbar}{2} i \langle S_l \rangle. \tag{15}$$

Curiously, one can write a similar relation for the full spins with equal phase angles $\varphi(j) = \varphi(k)$, i.e.:

$$\tfrac{1}{2}\langle [S_j, S_k] \rangle_{\varphi(l)} \equiv \tfrac{1}{2}R_{\varphi(l)}\langle [S_j, S_k] \rangle R^\dagger_{\varphi(l)} = \varepsilon_{jkl} \tfrac{\hbar}{2} i S_l. \tag{16}$$

Similarly, it is straightforward to check that the corresponding anti-commutators are zero, as expected:

$$\{\langle S_j \rangle, \langle S_k \rangle\} = 0; \qquad \langle \{S_j, S_k\} \rangle_{\varphi(l)} = 0 \tag{17}$$

What are the relations corresponding to the *eigen*-values for the spin basis in (8)? It is easy to check that:

$$\langle \boldsymbol{\sigma}_j S_3 \rangle = \begin{cases} 0 & \text{for } j = 1,2 \\ \pm \hbar/2 & \text{for } j = 3 \end{cases} \tag{18}$$

From Eq. (8), $S_\sigma \equiv S_3$ stands for either spin on the $\boldsymbol{\sigma}_3$ axis, which yields the (normalized) expectation values $\pm 1$ in the $\boldsymbol{\sigma}_3$ direction. The zero expectation value for $j = 1,2$ expresses the experimental finding of equal probabilities for spin up and spin down when $S_3$ is tested in directions perpendicular to $\boldsymbol{\sigma}_3$. Equation (18) is a special case of the general expectation value for $S_3$ after a *SG* transformation onto the direction **u**. Returning to the default notation $S_3 \equiv S_\sigma$ this is expressed either by the correlation between the start and end spins as in (19) below, or by the projection of the start spin onto the *SG* detector alignment **u** as in (20) further down:

$$\tfrac{4}{\hbar^2}\langle S_\sigma S_\mathbf{u}\rangle_{(SG)} = \tfrac{4}{\hbar^2}\langle\langle S_\sigma\rangle\langle S_\mathbf{u}\rangle\rangle = \langle\boldsymbol{\sigma}_3\mathbf{u}\rangle_0 = \boldsymbol{\sigma}_3\cdot\mathbf{u} = (-\boldsymbol{\sigma}_3)\cdot(-\mathbf{u}) = \cos\theta_\mathbf{u} = \cos^2\tfrac{\theta_\mathbf{u}}{2} - \sin^2\tfrac{\theta_\mathbf{u}}{2}, \quad (19)$$

$$\tfrac{2}{\hbar}\langle S_\sigma\mathbf{u}\rangle_{(SG)} = \langle\boldsymbol{\sigma}_3\mathbf{u}\rangle_0 = \langle\mathbf{u}\boldsymbol{\sigma}_3\rangle_0 = \cos\theta_\mathbf{u} = \cos^2\tfrac{\theta_\mathbf{u}}{2} - \sin^2\tfrac{\theta_\mathbf{u}}{2} \quad (20)$$

The *probabilities* for coincidence and anticoincidence outcomes are $\cos^2\theta_\mathbf{u}/2$ and $\sin^2\theta_\mathbf{u}/2$, which is expected remembering the probability amplitudes in (12). (20) tells us that the detector alignment conditions the final spin axis. For later reference, note that (19, 20) yield the expectation value of one spin measurement by projecting out the wedge part of $\boldsymbol{\sigma}_3\mathbf{u} = \boldsymbol{\sigma}_3\cdot\mathbf{u} + \boldsymbol{\sigma}_3\wedge\mathbf{u}$. For measurements on entangled pairs, the scalar parts still determine the partial expectation value for one-particle observations (though taking account of superposition), but *pair* correlations comprise contribution from *both* the scalar and the bivector parts.

Now, if the relative probabilities and probability amplitudes show up in relations involving only spin vectors, like (12), why do we need the full spin (8)? Well, the full spin connects to the total angular momentum of spin ½ and hints to additional spin structure compared to the measured spin, ultimately grounding onto the uncertainty principle. In this sense, it is analogue to the standard Pauli spin vector operator. In addition, the explicit gauge phase formalizes the entrance of irreversibility with measurement due to the complete loss of phase correlation, thus resulting, as shown in (11), into the restricted transformations (12). In the single-spin case, one can pick $\{\ell\}$ or $\{r\}$ handedness, while in the two-spin case we need both. *Phase* and *handedness* are key concepts for understanding the intrinsic total spin angular momentum of entangled pairs. Again, as in the one-spin case, the parts with phase contribute by zero expectation values in measurement correlations. In order to render the formulation as clear as possible, I use an arrow $\xrightarrow{SG}$, or $\rightarrow$ *only* for spin measurements.

**3.2. Two-particle systems.** The above discussion for the one-particle case generalizes in the following way to two-particle systems, again preserving a clear geometric picture of spin. The total angular momentum of *two spins* is the vector sum of the single total angular momenta, i.e. (see (2) for the definition of scalar product):

$$S_{tot} = S_{(1)} + S_{(2)}; \quad S_{tot}^2 = S_{(1)}^2 + S_{(2)}^2 + S_{(1)}S_{(2)} + S_{(2)}S_{(1)} = \tfrac{3\hbar^2}{2} + 2S_{(1)}\cdot S_{(2)} \quad (21)$$

$S_{tot}$ is Hermitian and symmetric relative to the swap of the two spins. The four *maximally* entangled (Bell) states [11, 13] consist of the superposition, shown by the *swap* sign ⇆, of two 2-spin states with the spins in each state being *specular* or *inverse* images of each other. In terms of the basis spins $S_\sigma$ they read:

$$Y_{(\mu)}: (S_{(1)}, S_{(2)})_{(\mu)} \equiv \{((S_\sigma)_\varphi, (i\sigma_\mu S_\sigma \sigma_\mu i)_\varphi) \leftrightarrows ((i\sigma_\mu S_\sigma \sigma_\mu i)_\varphi, (S_\sigma)_\varphi)\}; \quad \mu = 0,1,2,3 \quad (22)$$

The meaning of Eq. (22) is that the state $Y_{(\mu)}$ weighs equally the two 2-spin states at each side of the swap sign ⇆, thus capturing the essence of superposition of 2-spin states. As an illustration, in Eq. (23) below we will calculate the *intrinsic* total spin $S_{tot(\mu)}$ for the states $Y_{(\mu)}$. Due to the superposition in (22), $S_{tot(\mu)} = \frac{1}{2}[(S_\sigma + i\sigma_\mu S_\sigma \sigma_\mu i)_{(1)} + (i\sigma_\mu S_\sigma \sigma_\mu i + S_\sigma)_{(2)}] = S_\sigma + i\sigma_\mu S_\sigma \sigma_\mu i$ (phase not shown for simplicity); the last expression appears in Eq. (23). One could have tried a form for $Y_{(\mu)}$ reminiscent to that of the standard definition; see the alternative form $Y^{(alt)}_{SG(\mu)}$ in Eq. (27) and the *SG* expectation value in Eq. (28). We calculate now the intrinsic $S_{tot}$, $S^2_{tot}$, $S_{(1)} \cdot S_{(2)}$ from Eq. (21) (below $\sigma_j S_\sigma \sigma_j = -S_\sigma + 2(S_\sigma \cdot \sigma_j)\sigma_j$, no sum on *j*):

$$S_{tot(\mu)} = S_{(1)} + S_{(2)} = (S_\sigma - \sigma_\mu S_\sigma \sigma_\mu)_\varphi = \begin{cases} 0; & (S_{tot(0)})^2 = 0; & Y_{(0)} \\ 2[S_\sigma - (S_\sigma \cdot \sigma_j)\sigma_j]; & (S_{tot(j)})^2 = 2\hbar^2; & Y_{(j)} \end{cases} \quad (23)$$

$$2(S_{(1)} \cdot S_{(2)})_{(\mu)} = -2(S_\sigma)_\varphi \cdot (\sigma_\mu S_\sigma \sigma_\mu)_\varphi = \begin{cases} -2S^2 = -3\hbar^2/2; & Y_{(0)} \\ 2[3\hbar^2/4 - \hbar^2/2] = \hbar^2/2; & Y_{(j)}(j = 1,2,3) \end{cases} \quad (24)$$

From (22) the maximally entangled spins have *identical* phase; they relate by inversion in the singlet state $Y_{(0)}$ and by plane-reflections in the three triplet states $Y_{(j)}$. As already mentioned, the vector model 'places' both spins before observation in the same 3D orientation space, avoiding the tensor product and interpreting, e.g. the zero total spin for $Y_{(0)}$ in (23) as due to two opposite spins. The expression for the total angular momentum $S^2 = \hbar^2 S(S + 1)$ and Eq. (23) tell us that $S = 1$ (or 0) – i.e. an *observed* combined spin of modulus $\sqrt{\langle S \rangle^2} = \hbar$ (or 0) in the triplet (resp. singlet) states. Looking instead at $\langle S_{tot} \rangle^2 = (\langle S_{(1)} \rangle + \langle S_{(2)} \rangle)^2$ we find $\hbar^2$ (or 0) for $Y_{(1,2)}$ (resp. $Y_{(0,3)}$), corresponding to the standard eigenvalues $\pm\hbar$ (resp. 0) along $\sigma_3$. From the above discussion and following the convention [11, 13], we further define $Y_{(1)}, Y_{(2)}$ to comprise two spins up, respectively two spins down. One choice of mutually *orthogonal* states, as demonstrated in the spinor form in Section 4.2, is the following (the phase $\varphi$, which as in (22) is the same for all spins below, is not shown for simplicity):

$$\frac{\hbar}{2}\begin{cases}((\sigma_3+\sigma_1+\sigma_2)_{(1)},(-\sigma_3-\sigma_1-\sigma_2)_{(2)}) \leftrightarrows ((-\sigma_3-\sigma_1-\sigma_2)_{(1)},(\sigma_3+\sigma_1+\sigma_2)_{(2)}); & \Upsilon_{(0)} \\ ((\sigma_3+\sigma_1+\sigma_2)_{(1)},(\sigma_3-\sigma_1+\sigma_2)_{(2)}) \leftrightarrows ((\sigma_3-\sigma_1+\sigma_2)_{(1)},(\sigma_3+\sigma_1+\sigma_2)_{(2)}); & \Upsilon_{(1)} \\ ((-\sigma_3-\sigma_1-\sigma_2)_{(1)},(-\sigma_3-\sigma_1+\sigma_2)_{(2)}) \leftrightarrows ((-\sigma_3-\sigma_1+\sigma_2)_{(1)},(-\sigma_3-\sigma_1-\sigma_2)_{(2)}); & \Upsilon_{(2)} \\ ((-\sigma_3-\sigma_1-\sigma_2)_{(1)},(\sigma_3-\sigma_1-\sigma_2)_{(2)}) \leftrightarrows ((\sigma_3-\sigma_1-\sigma_2)_{(1)},(-\sigma_3-\sigma_1-\sigma_2)_{(2)}); & \Upsilon_{(3)}\end{cases} \quad (25)$$

In (25) the spin sums for $\Upsilon_{(1)}, \Upsilon_{(2)}$ lie on the respective cones defined by $\sigma_3 + (\sigma_2)_\varphi$ and $-\sigma_3 - (\sigma_1)_\varphi$, meeting at the origin and with a phase difference of $\pi/2$; while the spin sum for $\Upsilon_{(3)}$ lies in the plane $-(\sigma_1 + \sigma_2)_\varphi$. Within phase factors, these agree with the $S_{tot(\mu)}$ in (23).

**SG transformations.** Recalling now the generic form of the observed spin in Eq. (12), $\frac{\hbar}{2} R_w \sigma_3 R_w^\dagger$ (slightly changed notation), we can depict the four Bell states for bipartite observations with two *SG* magnets aligned along $\mathbf{w} = \mathbf{u}, \mathbf{v}$ and with superposition ($\leftrightarrows$) as:

$$\Upsilon_{SG(\mu)} \equiv \frac{\hbar}{2}\{(R_u\sigma_3 R_u^\dagger, i\sigma_\mu R_v\sigma_3 R_v^\dagger i\sigma_\mu) \leftrightarrows (i\sigma_\mu R_u\sigma_3 R_u^\dagger i\sigma_\mu, R_v\sigma_3 R_v^\dagger)\} =$$

$$\frac{\hbar}{2}\{(\mathbf{u}, i\sigma_\mu \mathbf{v} i\sigma_\mu) \leftrightarrows (i\sigma_\mu \mathbf{u} i\sigma_\mu, \mathbf{v})\}. \tag{26}$$

An *alt*ernative form of entangled states that fits the calculation of the expectation value in (28) is:

$$\Upsilon_{SG(\mu)}^{(alt)} \equiv \frac{\hbar^2}{4}\left[(R_u\sigma_3 R_u^\dagger)_{(1)}(i\sigma_\mu R_v\sigma_3 R_v^\dagger i\sigma_\mu)_{(2)} + (i\sigma_\mu R_u\sigma_3 R_u^\dagger i\sigma_\mu)_{(1)}(R_v\sigma_3 R_v^\dagger)_{(2)}\right] \tag{27}$$

I will keep the form (26) in the following as it depicts the two-spin states by *spin pairs* (with superposition). Superposition is also a necessary condition for calculating bipartite correlations under a *common* angled bracket. The respective *normalized* expectation values with superposition yield the standard results:

$$\langle\Upsilon_{SG(\mu)}^{(alt)}\rangle = \langle\Upsilon_{SG(\mu)}\rangle \equiv \tfrac{1}{2}\langle \mathbf{u} i\sigma_\mu \mathbf{v} i\sigma_\mu + i\sigma_\mu \mathbf{u} i\sigma_\mu \mathbf{v}\rangle = -\tfrac{1}{2}\langle \mathbf{u}\sigma_\mu \mathbf{v}\sigma_\mu + \sigma_\mu \mathbf{u}\sigma_\mu \mathbf{v}\rangle = -\langle \mathbf{u}\sigma_\mu \mathbf{v}\sigma_\mu\rangle = -\langle\sigma_\mu \mathbf{u}\sigma_\mu \mathbf{v}\rangle =$$

$$\begin{cases}-(\mathbf{u}\cdot\mathbf{v}) = -\cos\vartheta_{uv} = \sin^2\frac{\vartheta_{uv}}{2} - \cos^2\frac{\vartheta_{uv}}{2}; & \mu = 0 \\ \mathbf{u}\cdot\mathbf{v} - 2(\mathbf{u}\cdot\sigma_j)(\mathbf{v}\cdot\sigma_j) = \mathbf{u}\cdot\mathbf{v} - 2u^j v^j = \cos\vartheta_{uv} - 2\cos\vartheta_{uj}\cos\vartheta_{vj}; & \mu = j = 1,2,3\end{cases} \tag{28}$$

$u^k, v^k$ are the scalar components of $\mathbf{u}$ and $\mathbf{v}$ along $\sigma_k$, i.e. the expectation values for the triplet states depend on the reference frame. Both expectations for full spin in (24) and observed spin in (28) satisfy the '*closure*' relation: $\sum_\mu (S_{(1)} \cdot S_{(2)})_{(\mu)} = 0$, which is frame-independent. Each superposition term in (28) contributes equally to the expectation value; therefore, we could use one of them as shown by the third and fourth equalities in (28). Looking e.g. at the second term, the associativity of the geometric product in (1) allows to group the terms more symmetrically by splitting the improper rotation between the two detector alignments $(i\boldsymbol{\sigma}_\mu \mathbf{u} i\boldsymbol{\sigma}_\mu)\mathbf{v} = (i\boldsymbol{\sigma}_\mu \mathbf{u})(i\boldsymbol{\sigma}_\mu \mathbf{v})$. The last form hints to a spinor representation of the four possible states, as applied and discussed in Section 4.2, Eqs. (40, 41). While one of the superposition terms is sufficient to calculate the correct values for the bipartite expectations in (28), we will see shortly that both are needed to calculate the *partial* one-spin values for each of the entangled particles. Before doing that calculation, notice that the same bipartite expectation values as in (28) follow from the forms below, as alternatives to (26):

$$Y'_{SG(\mu)} \equiv \tfrac{\hbar}{2}\{(R_u(\pm\boldsymbol{\sigma}_3)R_u^\dagger, i\boldsymbol{\sigma}_\mu R_v(\pm\boldsymbol{\sigma}_3)R_v^\dagger i\boldsymbol{\sigma}_\mu) \leftrightarrows (i\boldsymbol{\sigma}_\mu R_u(\pm\boldsymbol{\sigma}_3)R_u^\dagger i\boldsymbol{\sigma}_\mu, R_v(\pm\boldsymbol{\sigma}_3)R_v^\dagger)\};$$

$$Y''_{SG(\mu=1,2)} \equiv \tfrac{\hbar}{2}\{(R_u(\pm\boldsymbol{\sigma}_3)R_u^\dagger, i\boldsymbol{\sigma}_\mu R_v(\pm\boldsymbol{\sigma}_3)R_v^\dagger i\boldsymbol{\sigma}_\mu) \leftrightarrows (i\boldsymbol{\sigma}_\mu R_u(\mp\boldsymbol{\sigma}_3)R_u^\dagger i\boldsymbol{\sigma}_\mu, R_v(\mp\boldsymbol{\sigma}_3)R_v^\dagger)\}. \quad (29)$$

The forms $Y'_{SG(\mu)}$ for $\mu = 0,1$ (upper signs) and $\mu = 2,3$ (lower signs) connect to the $\langle Y_{(\mu)}\rangle$ in (25), while the forms $Y''_{SG(\mu=1,2)}$ (applying only to $\mu = 1,2$, as $Y'_{SG(\mu=0,3)}$ correspond already to the standard $\Psi^\pm$) connect to the standard form most in use today: $\Phi^\pm = \tfrac{1}{\sqrt{2}}(|\uparrow\uparrow\rangle \pm |\downarrow\downarrow\rangle)$ (two spins up superposed to two spins down with opposite phases for the two triplet states). Of course, there is no tensor product in the present formalism!

We calculate now the *partial* one-spin expectation values for the four maximally entangled pairs in (26, 29). We will use (29) only for $\mu = 1,2$, as $Y'_{SG(\mu=0,3)}$ in (28) yields the same result as $Y_{SG(\mu=0,3)}$ in (26) without sign complications. The one-spin relation (19) applied to each particle becomes (now with superposition):

Spin(1): $\quad \tfrac{1}{2}\epsilon_{1,2}\langle\boldsymbol{\sigma}_3(\mathbf{u} + i\boldsymbol{\sigma}_\mu \mathbf{u} i\boldsymbol{\sigma}_\mu)\rangle = \tfrac{1}{2}\epsilon_{1,2}\boldsymbol{\sigma}_3 \cdot (\mathbf{u} - \boldsymbol{\sigma}_\mu \mathbf{u}\boldsymbol{\sigma}_\mu),\quad$ or equivalently

$\qquad\qquad \tfrac{1}{2}\epsilon_{1,2}\langle\mathbf{u}(\boldsymbol{\sigma}_3 + i\boldsymbol{\sigma}_\mu \boldsymbol{\sigma}_3 i\boldsymbol{\sigma}_\mu)\rangle = \tfrac{1}{2}\epsilon_{1,2}\mathbf{u} \cdot (\boldsymbol{\sigma}_3 - \boldsymbol{\sigma}_\mu \boldsymbol{\sigma}_3 \boldsymbol{\sigma}_\mu);$

Spin(2): $\quad \tfrac{1}{2}\epsilon_{1,2}\langle\boldsymbol{\sigma}_3(i\boldsymbol{\sigma}_\mu \mathbf{v} i\boldsymbol{\sigma}_\mu + \mathbf{v})\rangle = \tfrac{1}{2}\epsilon_{1,2}\boldsymbol{\sigma}_3 \cdot (\mathbf{v} - \boldsymbol{\sigma}_\mu \mathbf{v}\boldsymbol{\sigma}_\mu),\quad$ or equivalently

$\qquad\qquad \tfrac{1}{2}\epsilon_{1,2}\langle\mathbf{v}(\boldsymbol{\sigma}_3 + i\boldsymbol{\sigma}_\mu \boldsymbol{\sigma}_3 i\boldsymbol{\sigma}_\mu)\rangle = \tfrac{1}{2}\epsilon_{1,2}\mathbf{v} \cdot (\boldsymbol{\sigma}_3 - \boldsymbol{\sigma}_\mu \boldsymbol{\sigma}_3 \boldsymbol{\sigma}_\mu). \quad (30)$

It is clear from (30) that the states are not pure, as expected for partial states from entanglement. The partial expectation values belonging to $\mu = 0,3$, i.e. $Y_{(0)}, Y_{(3)}$ are zero. The factors $\epsilon_{1,2}$ apply only to the two cases of $Y_{(\mu=1,2)}$ and are $\epsilon_1 = 1, \epsilon_2 = -1$, reflecting the choice of spin vectors in $Y'_{SG(\mu=1,2)}$ of (29). With that choice the partial expectation values for $Y_{(\mu=1,2)}$ in (30) are $\epsilon_{1,2}u^3$ and $\epsilon_{1,2}v^3$, respectively. With the form $Y''_{SG(\mu)}$ instead of $Y'_{SG(\mu)}$ the two partial expectation values vanish; e.g., the value for Spin(1) would become $\tfrac{1}{2}\mathbf{u} \cdot (\epsilon_{1,2}\boldsymbol{\sigma}_3 - \epsilon_{2,1}\boldsymbol{\sigma}_\mu \boldsymbol{\sigma}_3 \boldsymbol{\sigma}_\mu) = 0$ ($\mu = 1,2$) (the first (second) subscripts in $\epsilon_{1,2}, \epsilon_{2,1}$ apply to $Y''_{SG(\mu=1)}$ (resp. $Y''_{SG(\mu=2)}$)). With this choice, all the partial expectation values for the four Bell states *vanish*.

The bivector form of the partial expectation values in (30) can also serve as an *alternative* starting point to obtain the *bipartite* expectation values in (28) (bivectors like $\sigma_3 \mathbf{u}$ are not Hermitian), i.e.:

Spins(1)&(2): $\langle \Upsilon_{SG(\mu)} \rangle = \langle (\sigma_3 \mathbf{u})^\dagger \sigma_3 \mathbf{i} \sigma_\mu \mathbf{v} \mathbf{i} \sigma_\mu \rangle = -\langle \mathbf{u} \sigma_3 \sigma_3 \sigma_\mu \mathbf{v} \sigma_\mu \rangle = -\langle \mathbf{u} \sigma_\mu \mathbf{v} \sigma_\mu \rangle$, or equivalently

$$\langle \Upsilon_{SG(\mu)} \rangle = \langle (\sigma_3 \mathbf{i} \sigma_\mu \mathbf{u} \mathbf{i} \sigma_\mu)^\dagger \sigma_3 \mathbf{v} \rangle = -\langle \sigma_\mu \mathbf{u} \sigma_\mu \mathbf{v} \rangle. \tag{31}$$

This form proves the statement following Eq. (20) that both the *scalar* and the *bivector* parts of the geometric products $\sigma_3 \mathbf{u} = \sigma_3 \cdot \mathbf{u} + \sigma_3 \wedge \mathbf{u}$ and $\sigma_3 \mathbf{i} \sigma_\mu \mathbf{v} \mathbf{i} \sigma_\mu = -\sigma_3 \cdot (\sigma_\mu \mathbf{v} \sigma_\mu) - \sigma_3 \wedge (\sigma_\mu \mathbf{v} \sigma_\mu)$ contribute to the entanglement correlations. As we will see in Eq. (32) below, this distinguishes entangled pairs from separable, non-entangled pairs where the bivector parts do not contribute to the bipartite correlations.

How do *separable* states look like in the present formalism? Superposition in this case lose the redundancy we saw in the case of entanglement, i.e. the equal contribution from the two terms in (28, 31). Each separable state can be expressed by cross-superposition, as shown below (first raw combines pieces from $\Upsilon_{(0)}, \Upsilon_{(3)}$ (changed ordering yields either ↑↓ (shown) or ↓↑); second raw combines pieces from $\Upsilon_{(1)}, \Upsilon_{(2)}$ (changing signs as in (29) it yields either ↑↑ (shown) or ↓↓)), where we also show the expectation values:

$\frac{\hbar}{2}\{(\mathbf{u}, \mathbf{i}\sigma_0 \mathbf{v} \mathbf{i} \sigma_0) \leftrightarrows (-\mathbf{i}\sigma_3 \mathbf{u} \mathbf{i} \sigma_3, -\mathbf{v})\} \Rightarrow \frac{1}{2}\langle -\mathbf{u}\mathbf{v} - \sigma_3 \mathbf{u} \sigma_3 \mathbf{v} \rangle = \frac{1}{2}\langle -\mathbf{u}\mathbf{v} + \mathbf{u}\mathbf{v} - 2u^3 v^3 \rangle = -u^3 v^3;$

$\frac{\hbar}{2}\{(\mathbf{u}, \mathbf{i}\sigma_1 \mathbf{v} \mathbf{i} \sigma_1) \leftrightarrows (\mathbf{i}\sigma_2 \mathbf{u} \mathbf{i} \sigma_2, \mathbf{v})\} \Rightarrow \frac{1}{2}\langle -\mathbf{u}\sigma_1 \mathbf{v} \sigma_1 - \sigma_2 \mathbf{u} \sigma_2 \mathbf{v} \rangle = \frac{1}{2}\langle \mathbf{u}\mathbf{v} + \mathbf{u}\mathbf{v} - 2u^1 v^1 - 2u^2 v^2 \rangle = u^3 v^3. \tag{32}$

The bipartite expectation value $-u^3 v^3$ is the same for antiparallel spins, ↑↓ and ↓↑, while $u^3 v^3$ is valid for parallel spins, ↑↑ and ↓↓. It is easy to check that the partial expectation values for the four separable states ordered in pairs are $(u^3, -v^3); (-u^3, v^3); (u^3, v^3); (-u^3, -v^3)$. The products for each pair reproduce the two bipartite expectation values in (32) – a feature of the separable states.

Even more clearly, the separable states show exactly the same correlations, bipartite and partial, as pairs of (non-entangled) pure one-spin states, thus justifying the designation 'separable'. For example, the bipartite expectation value for antiparallel spins, each expressed as in Eq. (20) reproduces the top-line result in (32):

$$\langle S_{(1)} S_{(2)} \rangle_0 = \frac{4}{\hbar^2} \langle\langle S_\sigma \rangle \mathbf{u} \rangle \langle \langle S_{-\sigma} \rangle \mathbf{v} \rangle = -(\sigma_3 \cdot \mathbf{u})(\sigma_3 \cdot \mathbf{v}) = -u^3 v^3 = -\cos\theta_\mathbf{u} \cos\theta_\mathbf{v}. \tag{33}$$

What about the *orthogonality* of the maximally entangled states? Like the one-particle case (remember $\mathbf{u}^+, \mathbf{u}^-$ in Fig. 1b), orthogonality is experienced in the spinor (one-sided rotor) representation, as described by the Equations (42-44) and the related discussions in Section 4.2.

## 4. Spinor (one-sided rotor) representation of spin 1/2 one and two particle systems.

In spinor form, the transformation of a given spin appears as a *sum* (see (9)) of one-sided rotor operations of spin-up and spin-down with the standard probability amplitudes for coincidence and anti-coincidence.

**4.1. *One-particle systems.*** Instead of the two-sided rotor vector expression (10) we now express a given spin $S_{u(\varphi)}$ as a *sum* (+ sign, see (9) and ensuing discussion) of the (one-sided rotor) *spinor* forms for the spin basis, as illustrated in Fig. 1b (remember that the orientation of the angle $\theta_u$ is as seen from $\mathbf{u}_2$):

$$S_{u(\varphi-\varphi_u)} = R_{\theta_u} S_{\sigma(\varphi)} \cos\frac{\theta_u}{2} + R^\dagger_{\pi-\theta_u} S_{-\sigma(-\varphi)} \sin\frac{\theta_u}{2} = R_{\theta_u} S_{\sigma(\varphi)} \cos\frac{\theta_u}{2} + R^\dagger_{(\pi-\theta)_u} R_{\pi_u} S_{\sigma(\varphi)} R^\dagger_{\pi_u} \sin\frac{\theta_u}{2} =$$

$$R_{\theta_u} S_{\sigma(\varphi)} \left( \cos\frac{\theta_u}{2} + \mathbf{i}\mathbf{u}_2 \sin\frac{\theta_u}{2} \right) = R_{\theta_u} S_{\sigma(\varphi)} R^\dagger_{\theta_u} , \text{ or} \quad (34)$$

$$S_{u(\varphi-\varphi_u)} = \frac{\hbar}{2}\left[ (\mathbf{u}^+ + R_{\theta_u}(\sigma_1+\sigma_2)_\varphi) \cos\frac{\theta_u}{2} + (\mathbf{u}^- - R^\dagger_{\pi-\theta_u}(\sigma_1-\sigma_2)_{-\varphi}) \sin\frac{\theta_u}{2} \right] =$$

$$\frac{\hbar}{2}\left( \mathbf{u}^+ \cos\frac{\theta_u}{2} + \mathbf{u}^- \sin\frac{\theta_u}{2} + R_{\theta_u}\left( (\mathbf{u}_\perp+\mathbf{u}_2)_{\varphi-\varphi_u} \cos\frac{\theta_u}{2} - \mathbf{i}\mathbf{u}_2 (\mathbf{u}_\perp-\mathbf{u}_2)_{-\varphi+\varphi_u} \sin\frac{\theta_u}{2} \right) \right) =$$

$$\frac{\hbar}{2}\left( \mathbf{u} + R_{\theta_u}(\mathbf{u}_\perp+\mathbf{u}_2)_{\varphi-\varphi_u} \left( \cos\frac{\theta_u}{2} + \mathbf{i}\mathbf{u}_2 \sin\frac{\theta_u}{2} \right) \right) = \frac{\hbar}{2}\left( \mathbf{u} + R_{\theta_u}(\mathbf{u}_\perp+\mathbf{u}_2)_{\varphi-\varphi_u} R^\dagger_{\theta_u} \right) =$$

$$\frac{\hbar}{2}\left( \mathbf{u} + (\mathbf{u}_1+\mathbf{u}_2)_{\varphi-\varphi_u} \right). \quad (35)$$

By insisting that a basis consist of *opposite* spins then in order for (34, 35) to result in a proper rotation as they do, the two spins must have the same handedness. The plane for the phase angles $\varphi - \varphi_u$ in the subscripts is the plane defined by the two vectors $\mathbf{u}_1, \mathbf{u}_2$ in brackets. The concise form (34) is typical for STR/STA, where full geometric objects transform as a whole, in contrast to the form (35) where components do appear. Keeping the gauge phase implicit, the full *spinor representations* for spin up and spin down are:

$$R_{\theta_u} S_\sigma = \mathbf{u}^+ + R_{\theta_u}(\mathbf{u}_\perp+\mathbf{u}_2) \quad \text{and} \quad \mathbf{i}\mathbf{u}_2 R_{\theta_u} S_{-\sigma} = \mathbf{u}^- + R_{\theta_u}(\mathbf{u}_\perp+\mathbf{u}_2)\mathbf{i}\mathbf{u}_2 \quad (36)$$

Spinor representations *are not* spin according to Definition (8). As already mentioned, the midway vectors $\mathbf{u}^+$ and $\mathbf{u}^-$ are *reduced* spinor representations of the two spins; they are manifestly orthonormal. Spinor representations are not unique. As anticipated in Eq. (13), another reduced spinor representation comprises the factors in front of the rotor $R_{\theta_u}$ in (36), a scalar 1 and a bivector $\mathbf{i}\mathbf{u}_2$, respectively, which are also part of $R_{\theta_u}$. These are even-grade elements of the 3D algebra and a standard choice in the STA literature [7], see the discussion of Eq. (13). In order for the reduced representation to be an orthonormal spinor basis one must have a zero scalar (grade 0) for $\langle 1(\mathbf{i}\mathbf{u}_2)\rangle_0 = \langle \mathbf{u}^+\mathbf{u}^-\rangle_0 = 0$, which is indeed the case. The orthogonality relation for the full spinor representation is (remember that $S_\sigma$ is Hermitian):

$$\langle (R_{\theta_u}S_\sigma \mathbf{i}\mathbf{u}_2)^\dagger (R_{\theta_u}S_\sigma)\rangle_0 = -\langle \tfrac{3\hbar^2}{4}\mathbf{i}\mathbf{u}_2\rangle_0 = 0, \tag{37}$$

which is clearly satisfied. The orthogonality condition for the reduced representation is the normalized (37).

*Measurement.* The action of a *SG* magnet aligned along $\boldsymbol{\sigma}_3$ on a spin with original spin vector $\tfrac{\hbar}{2}\mathbf{u} = \tfrac{\hbar}{2}\mathbf{u}_3$ corresponds to an irreversible transformation relative to the gauge phase and instead of (34) we get:

$$S_{\mathbf{u}(\varphi)} \xrightarrow{SG} R_{\theta_u}^\dagger S_{\mathbf{u}(\varphi\prime)} = S_{\sigma(\varphi\prime)}\cos\tfrac{\theta_u}{2} + \mathbf{i}\mathbf{u}_2 S_{-\sigma(-\varphi\prime)}\sin\tfrac{\theta_u}{2}; \quad \langle \varphi\varphi'\rangle = \langle \varphi\rangle\langle \varphi'\rangle \tag{38}$$

In the *SG* case it is straightforward to apply the spinor transformation to the spin vector alone, see (35):

$$\tfrac{\hbar}{2}\mathbf{u} \xrightarrow{SG} \tfrac{\hbar}{2}R_{\theta_u}^\dagger \mathbf{u} = \tfrac{\hbar}{2}\boldsymbol{\sigma}_3\left(\cos\tfrac{\theta_u}{2} + \mathbf{i}\mathbf{u}_2\sin\tfrac{\theta_u}{2}\right) = \tfrac{\hbar}{2}R_{\theta_u}^\dagger\left(\mathbf{u}^+\cos\tfrac{\theta_u}{2} + \mathbf{u}^-\sin\tfrac{\theta_u}{2}\right) \tag{39}$$

**4.2. *Two-particle systems.*** The scope of this Subsection is to prove the mutual orthogonality among the four Bell states. The proof will equally apply to the separable states in (32), as they consist of the same pairs as the Bell states, just combined differently. Let us start by writing the non-normalized expression in (28) for $\mathbf{u} = \mathbf{v} = \boldsymbol{\sigma}_3$ as (notice that $S_\sigma \mathbf{i}\boldsymbol{\sigma}_\mu$ is not Hermitian):

$$\langle S_\sigma \mathbf{i}\boldsymbol{\sigma}_\mu S_\sigma \mathbf{i}\boldsymbol{\sigma}_\mu\rangle_0 = \langle (-\mathbf{i}\boldsymbol{\sigma}_\mu S_\sigma)^\dagger (S_\sigma \mathbf{i}\boldsymbol{\sigma}_\mu)\rangle_0. \tag{40}$$

Then one realizes that a non-normalized spinor form for the entangled states consists of the superposed pairs:

$$Y_{(s\mu)}: \{(-\mathbf{i}\boldsymbol{\sigma}_\mu S_\sigma, \mathbf{i}\boldsymbol{\sigma}_\mu S'_\sigma) \rightleftarrows (\mathbf{i}\boldsymbol{\sigma}_\mu S'_\sigma, -\mathbf{i}\boldsymbol{\sigma}_\mu S_\sigma)\} \quad \text{with} \quad \mathbf{i}\boldsymbol{\sigma}_\mu S'_\sigma \equiv \mathbf{i}\boldsymbol{\sigma}_\mu \boldsymbol{\sigma}_\mu S_\sigma \boldsymbol{\sigma}_\mu = S_\sigma \mathbf{i}\boldsymbol{\sigma}_\mu \tag{41}$$

The suffix s in $Y_{(s\mu)}$ stands for 'spinor representation'. As in Eqs. (22, 28, 31), superposition produces two equally contributing terms in the expressions for the bipartite expectation values. It is sufficient to prove orthogonality by looking at one of the pairs. Eq. (41) for the four states comprise terms that are either even (bivectors) for the singlet ($\mu = 0$) or odd (vectors and pseudoscalar) for the triplet ($\mu = j$) states, therefore:

$$\langle Y_{(0)}^\dagger Y_{(j)}\rangle_0 = \langle Y_{(j)}^\dagger Y_{(0)}\rangle_0 = 0 \tag{42}$$

for both full (intrinsic) and measured (extrinsic) spins. In words, the singlet state is orthogonal to the triplet states. By taking the full spins at the left (resp. right) in each pair in (25) as $S_\sigma$ ($S'_\sigma$ from (41)), one can prove mutual orthogonality for all pairs of states ($\mu \neq \nu$):

$$\langle \Upsilon^\dagger_{(\mu)} \Upsilon_{(\nu)} \rangle_0 = \langle S_\sigma \mathrm{i}\boldsymbol{\sigma}_\mu (\mathrm{i}\boldsymbol{\sigma}_\nu S'_\sigma) \rangle_0 = -\langle S_\sigma \boldsymbol{\sigma}_{\mu\nu} S'_\sigma \rangle_0 = \langle C S_\sigma \boldsymbol{\sigma}_l S'_\sigma \rangle_0 =$$

$$\pm \hbar^2/4 \, \langle C(\boldsymbol{\sigma}_3 + \boldsymbol{\sigma}_1 + \boldsymbol{\sigma}_2) \boldsymbol{\sigma}_l (\boldsymbol{\sigma}_3 + \boldsymbol{\sigma}_1 + \boldsymbol{\sigma}_2) \rangle_0 = 0;$$

$$C = 1 \text{ for } \mu\nu = 0; \quad C = \mathrm{i}\varepsilon_{jkl} \text{ for } \mu\nu = jk; \quad j,k,l = 1,2,3 \tag{43}$$

Finally, the orthogonality relations among the *SG* measured states are straightforward:

$$\langle \Upsilon^\dagger_{(\mu)} \Upsilon_{(\nu)} \rangle_{0(SG)} = \langle \boldsymbol{\sigma}_3 \mathrm{i}\boldsymbol{\sigma}_\mu (\mathrm{i}\boldsymbol{\sigma}_\nu \boldsymbol{\sigma}_\nu \boldsymbol{\sigma}_3 \boldsymbol{\sigma}_\nu) \rangle_0 = \pm \langle \boldsymbol{\sigma}_\mu \boldsymbol{\sigma}_\nu \rangle_0 = 0 \text{ for } \mu \neq \nu; \quad \mu,\nu = 0,1,2,3 \tag{44}$$

From (43, 44) both the full and the *SG* spinor representations for the maximally entangled states are *mutually orthogonal*. The last Equation in (44) essentially reduces the orthogonality relation for the four basis 2-spin states (maximally entangled or separable) into the orthogonality of the Pauli basis $\{\boldsymbol{\sigma}_\mu\} = \{1, \boldsymbol{\sigma}_j\}$! The same result is obtained by using the one-sided factors $\mathrm{i}\boldsymbol{\sigma}_\mu$ in front of $S_\sigma, S'_\sigma$ in (41). Each of these is 'half' of one of the operations of inversion and the three plane reflections characterizing the four entanglement states $\Upsilon_{(\mu)}$, $\mu = 0,1,2,3$. We do not need to calculate the bipartite expectation values for entangled pairs in spinor representation, as by construction they are equal to the results in Eq. (29).

A swift comparison of the two-particle spinor representation (41) with the vector representation (22) reveals that inversion and reflection apply to *one* of the spins in (22) as two-sided operations $\mathrm{i}\boldsymbol{\sigma}_\mu S_\sigma \mathrm{i}\boldsymbol{\sigma}_\mu$ (though with superposition!). In the spinor representation (41) the same improper rotations split between the two spins, 'half rotation' for each, thus appearing as one-sided operations, e.g. $\mathrm{i}\boldsymbol{\sigma}_\mu S'_\sigma$. Of course, superposition is also present in (41). This makes the states in (41) apparently more symmetric but also more abstract then the states in (22). However, I stress again that the improper rotations contribute to both spins in Eqs. (22, 26) due to superposition. In addition, the correlations in the form (28) are the same for both representations and as *correlation* expectation values they take account of both entangled spins under the *same* angled brackets. This would represent a nonlocal operation, as the two measurements actually take place at different locations. To be sure, the common angled brackets refer to the *statistical dependen*ce between the two measurements. Anyway, cutting a potentially long discussion short, it is relevant to remind the reader here that we are working in the spin-position decoupling approximation and in this framework, the operation of common angled brackets is *local*. Notice that the bipartite states respect Pauli's exclusion principle, as they had to at the pairs' creation.

**Conclusions**

The spin model in (8) – sum of three orthogonal vectors with a phase in the 3D orientation space, replaces the symbolic Pauli spin matrix-vector of standard QM. By the same move, proper and improper rotor operations on vectors substitute the standard operator-state formalism. The model displays many attractive features in the spin-position decoupling approximation. For one and two particle systems, it shows the correct representation of spin 1/2 relative to both *intrinsic* full spin and *observed* spin expectations. The explicit gauge phase in (8) allows formalizing the irreversibility related to measurement as due to loss of phase correlation. The adaptive embedding of the spin space illustrated in Fig. 1b is remarkable, proving the geometry of the one-spin basis in 3D. The 3D setting also offers a clear geometric meaning for the four Bell states: the entangled spins relate by the four basic improper rotations, by superposition and are in phase. All orthogonality relations apply in the 'spinor' representation and in the bipartite case ultimately reduce to the orthogonality of the Pauli basis. The results above were obtained by direct use and transformation (rotation, reflection) of spin(s) in the 3D orientation space, without invoking the *eigen*-algebra or the tensor product, which seem so fundamental in the standard formulation of the quantum spin.

**Appendix**

The STR approach is the shortest path from the relativistic 4-momentum to the Dirac Equation (STR DE). Alike STA, STR does not comprise matrices and all the complex structure arises from vectors and multivectors, not from scalars. However, unlike in STA DE, spin has not been put by hand in STR DE.

Keeping with tradition, I use in the following the reciprocal frame vectors $e^v$, which relate to the frame vectors in (3) by the STR signature $e_\tau = \zeta_{\tau v} e^v$. The STR pseudoscalar in the reciprocal basis is $\dot{I} = e^{01235}$. In coordinate representation, the spacetime frame vectors $e^\mu$ replace the standard Dirac matrices $\gamma^\mu$. All predictions of the standard DE (not manifestly covariant), follow from the manifestly covariant STR DE, thus rendering superfluous any preconceptions on 'internal degrees of freedom' of the electron allegedly represented by the standard $\gamma^\mu$ matrices [2]. STR DE affirms that the free Dirac electron / positron are the result of the *relativistic* 4-momentum and the postulate of *quantization*, nothing else.

Shortly, the quantization of the 4-momentum p of modulus $m$ yields the STR DE:

$$(\hat{p} - m)\psi = 0 \quad \text{with} \quad \hat{p} = \dot{\mathrm{i}}\hbar\nabla = \dot{\mathrm{i}}\hbar e^\mu \partial_\mu \equiv \dot{\mathrm{i}}\hbar e^\mu \, \partial/\partial x^\mu. \tag{A1}$$

The pseudoscalar $\dot{\mathrm{I}} = e^{01235}$ commutes with all elements of STR. The introduction of the Hermitian frame vector $e^5$ is *not* an additional postulate in STR; it is just a means to incorporate the quantization postulate, which in the standard approach arrives with the unit imaginary $i$, into a real vector space. Eq. (A1) is manifestly covariant. Both $\hat{p}$ and $m$ are relativistic invariants and under a Lorentz transformation $\mathcal{S}$ to the primed frame:

$$\{(\hat{p} - m)\psi = 0\} \to \{\mathcal{S}(\hat{p} - m)\psi = \mathcal{S}(\hat{p} - m)\mathcal{S}^{-1}\mathcal{S}\psi \equiv (\hat{p}' - m)\psi' = 0\}; \quad \hat{p}' = \hat{p}; \quad \psi' = \mathcal{S}\psi \tag{A2}$$

In the case of frame vectors $\mathcal{S} = S \equiv e^{(-\dot{\mathrm{i}}\sigma_j\vartheta_j + \mathbf{x}_j\alpha_j)/2}$ so that $e'^\mu = Se^\mu\tilde{S}$. S is obtained by exponentiation of boost vectors and rotor bivectors from Eqs. (4, 5); $\vartheta_j$ and $\alpha_j$ are Euclidean angles (see Eq. (6)) and rapidities (hyperbolic angles), respectively. Lorentz transformations and parity constrict the form of the spinor $\psi$ in (A3). For example, S 'bringing' elements from both **X** and **Σ** signals that $\psi$ belongs at least to the product space **XΣ**.

The STR spinor $\psi$ can be expressed with the help of two Pauli spinors $\varphi, e^5\chi$ (so that $(\varphi + \chi) \in \mathbf{X\Sigma}$) and of two orthogonal projectors $(1 \pm e^0)$ relative to the time axis $e^0$ (relating to parity [12]), i.e.:

$$\psi = \tfrac{1}{2}[(1 + e^0)\psi + (1 - e^0)\psi] = \varphi + \chi; \quad \varphi \equiv \tfrac{1}{2}(1 + e^0)\psi; \quad \chi \equiv \tfrac{1}{2}(1 - e^0)\psi; \quad \varphi, e^5\chi \in \mathbf{\Sigma} \tag{A3}$$

**Σ** is the subspace of axial vectors in STR (Eq. (5) in the main text) – isomorphic to the space of Pauli matrices.

Now, in the nonrelativistic regime, the dominant spinor $\varphi$ 'freed' from the fast oscillations yields the Pauli spinor proper $\varphi_P$ and the STR DE yields the Pauli Equation, STR PE. The STR Pauli Hamiltonian in the presence of an EM field takes the same form as the standard one (removing the hat from the operators):

$$\dot{\mathrm{i}}\hbar\partial_t\varphi_P = H_P\varphi_P = \left[\frac{\mathbf{P}^2}{2m} - eA_0 + \frac{\hbar e}{2m}(\boldsymbol{\sigma}, \mathbf{B})\right]\varphi_P \quad \text{with} \quad (\boldsymbol{\sigma}, \mathbf{B}) \equiv B_j\sigma_j \tag{A4}$$

The STR form of the Pauli spinor $\varphi_P$ in [12] is:

$$\varphi_P = \tfrac{1}{2}[(1 + \boldsymbol{\sigma}_3)\varphi_P + (1 - \boldsymbol{\sigma}_3)\varphi_P] \equiv \varphi_u + \boldsymbol{\sigma}_1\varphi_d;$$

$$\varphi_u \equiv \tfrac{1}{2}(1 + \boldsymbol{\sigma}_3)\varphi_P; \quad \boldsymbol{\sigma}_1\varphi_d \equiv \tfrac{1}{2}(1 - \boldsymbol{\sigma}_3)\varphi_P; \quad \varphi_u, \varphi_d \in \{a + \dot{\mathrm{i}}b; a, b \in \mathbb{R}\} \tag{A5}$$

$\varphi_u, \varphi_d$ are proportional to the probability amplitudes for spin up and spin down, respectively. Finally, in the spin-position decoupling approximation, *s-p*, i.e. for spin probability amplitudes not depending on position, one can factorize the common spatial dependence $\rho$ of $\varphi_u, \varphi_d$ and render the probability amplitudes explicit (the projector $(1 - \boldsymbol{\sigma}_3)$ in (A5) allows to trade $\boldsymbol{\sigma}_1\varphi_d$ with $-\dot{\mathrm{i}}\boldsymbol{\sigma}_2\varphi_d$ thus forming the rotor $R_\vartheta$ below):

$$\varphi_P \overset{s\text{-}p}{=} \rho R_\vartheta \equiv \rho e^{-\dot{\mathrm{i}}\boldsymbol{\sigma}_2\vartheta/2} = \rho\left(\cos\tfrac{\vartheta}{2} - \dot{\mathrm{i}}\boldsymbol{\sigma}_2 \sin\tfrac{\vartheta}{2}\right) \tag{A6}$$

The form of the Pauli spinor above is the same as in STA.

**Acknowledgment.** Thanks to the reviewers for their comments & critics of previous versions of the manuscript. This work does not have any conflicts of interest. There are no funders to report for this submission.**References**

[1] Pauli, W.: Zur Quantenmechanik des magnetischen Elektrons. Zeitschrift für Physik, 43 (9–10), 601–623 (1927).

[2] Dirac, P.A.M.: The quantum theory of the electron. Proc. Roy. Soc. Lon. A, 117:610 (1928).

[3] Grassmann, H.: Die Ausdehnungslehre. Enslin, Berlin (1862).

[4] Clifford, W.K.: Applications of Grassmann's extensive algebra. Am. J. Math., 1:350 (1878).

[5] Hestenes, D.: Space–Time Algebra. Gordon and Breach, New York (1966).

[6] Hestenes, D.: Oersted Medal Lecture 2002: Reforming the Mathematical Language of Physics. Am. J. Phys. 71, 104–121 (2003).

[7] Doran, C., Lasenby, A.: Geometric Algebra for Physicists, Cambridge University Press, Cambridge (2007).

[8] Doran, C.J.L., Lasenby, A.N. and Gull, S.F.: States and operators in the spacetime algebra. Found. Phys., 23(9):1239 (1993).

[9] Doran, C., Lasenby, A., Gull, S., Somaroo, S., Challinor, A.: Spacetime Algebra and Electron Physics. Adv. Imag. & Elect. Phys. 95, 271–365 (1996).

[10] Lasenby, A.: Geometric Algebra as a Unifying Language for Physics and Engineering and Its Use in the Study of Gravity. Adv. Appl. Clifford Algebras 27, 733–759 (2017).

[11] Bohm, D.: Quantum Theory (17.9), Prentice-Hall, Inc., Englewood Cliffs, New Jersey (1951), Ch. 17.

[12] Andoni, S.: Dirac Equation Redux by Direct Quantization of the 4-Momentum Vector, preprint posted on https://www.researchsquare.com/article/rs-313921/v8, DOI: 10.21203/rs.3.rs-313921/v8 (2022).

[13] Merzbacher, E.: Quantum Mechanics, J. Wiley & Sons, New York, Chichester, Weineheim, Brisbane, Singapore, Toronto (1998).
19